\documentclass[12pt]{extarticle}
\usepackage{amssymb,amsmath,amsfonts,amsthm,graphicx,psfrag}
\usepackage{cmap}
\usepackage{graphicx}
\usepackage[left=2.25cm,right=2.25cm,top=2.25cm,bottom=2.25cm]{geometry}
\usepackage{bm}
\usepackage{latexsym}

\title{The Relativistic Cornell-type Mechanism of Exotic Scalar Resonances}
\author{{\bf A.~M.~Badalian, M.~S.~Lukashov and Yu.~A.~Simonov} \\
 Institute  for Theoretical and Experimental Physics\\ (NRC ``Kurchatov Institute'')
 \\
B. Cheremushkinskaya 25, Moscow, 117218 Russia}
\date{\today\thanks{v3 (ed1)}}

\newcommand{\beq}{\begin{eqnarray}}
\newcommand{\eeq}{\end{eqnarray}}
\newcommand{\be}{\begin{equation}}
\newcommand{\ee}{\end{equation}}
\newcommand{\ben}{\begin{equation*}}
\newcommand{\een}{\end{equation*}}
\def\la{\mathrel{\mathpalette\fun <}}

\def\fun#1#2{\lower3.6pt\vbox{\baselineskip0pt\lineskip.9pt
\ialign{$\mathsurround=0pt#1\hfil ##\hfil$\crcr#2\crcr\sim\crcr}}}

\newcommand{{\SD}}{\rm SD}

\newcommand{\vex}{\mbox{\boldmath${\rm x}$}}
\newcommand{\vey}{\mbox{\boldmath${\rm y}$}}

\newcommand{\veP}{\mbox{\boldmath${\rm P}$}}
\newcommand{\vep}{\bm p}
\newcommand{\veq}{\mbox{\boldmath${\rm q}$}}

\newcommand{\veu}{\mbox{\boldmath${\rm u}$}}

 \newcommand{\oIm}{\operatorname{Im}}

 \newcommand{\oRe}{\operatorname{Re}}
\begin{document}
\maketitle
\begin{abstract}
The formalism of the coupled $q\bar q$ and the $\varphi\varphi  ( \pi-\pi$, $K\bar K, \pi K,...$) scalar  channels is  formulated, taking into account  the ground and radial  excited $q\bar q$ poles.  The basic role is shown to be played by the  transition coefficients  $k^{(I)} (q\bar q, |\varphi\varphi)$,  which are calculated using the quark-chiral Lagrangian without  free parameters. The resulting method, called the pole projection mechanism (PPM), ensures: 1) one resonance for each $\varphi\varphi$ channel from the basic $q\bar q$  pole, e.g. the $f_0 (500)$ resonance in the $\pi\pi$ channel;  2) a possibility to have two $\varphi\varphi$ resonances, coupled to the same $q\bar q$ state, when the channel coupling is taken into account in the meson-meson channels, which yields $f_0 (500)$ and  $f_0(980)$ from the same $n\bar n$ pole around 1 GeV; 3) the strong pole  shift  down for special ($\pi\pi, \pi K)$ channels  due to large transition coefficients  $k^{(I)}$, computed in this formalism without free parameters. The parameters of calculated complex poles  are  in   reasonable agreement with the experimental  data  of the resonances $f_0(500), f_0(980), a_0(980), a_0(1450), K^*_0(700), K^*_0(1430), f_0(1370), f_0(1710)$.

\end{abstract}
\maketitle

\section{Introduction}

The QCD theory of hadrons has very  developed resources to treat hadron properties and by now explained a majority of observed hadrons \cite{1}. Nevertheless, there exist hadronic objects, considered as non-standard  or extra states, with the properties (e.g. the masses and  widths) strongly  different from theoretical predictions \cite{2}, and most of them refer to light scalar  mesons, such as  $ f_0(500), f_0(980), a_0(980), K^*_0(700)$. They can  hardly be associated with the lowest conventional $q\bar q$
scalars for  several reasons: a) their masses are strongly displaced as compared to expected  $q\bar q$ masses; b) in some cases two  observed scalar resonances can be identified with one $q\bar q$ state with the same quantum numbers.

This situation is well described by Nils T\"{o}rnqvist in 1995 \cite{3}  ``Our present understanding of the light meson mass spectrum is in a  deplorable state... This is mainly because of the fact that... ``QCD inspired quark models'' fail so dramatically for scalar mesons...'' Nowadays, 25 years later, we have much better understanding of this topic. Indeed  continuous efforts of the physical community have brought a large amount of information about the properties of the scalars, their
decays, and production (see \cite{4,5,6,7,8,9,10} for reviews and analysis, and \cite{11,12,13} for most recent reviews). Theoretical approaches to the scalar spectrum include the tetraquark model \cite{14}, the chiral model \cite{15}, the molecular model \cite{16}, the QCD sum rules \cite{17}, and lattice calculations \cite{18}.  Our approach  is based on several premises:
 \begin{description}
 \item{1)} the primary poles are due to $q\bar q$ bound states, which are subject to interaction with meson-meson ($m-m$ systems);
 \item{2)} this interaction can be deduced from the quark-chiral Lagrangian without free parameters;
 \item{3)} the coupled channel interaction inside $m-m$ systems can connect more than one resonance to one original $q\bar q$ pole.
\end{description}
The similar ideas are not new and have been largely investigated since 1995 in \cite{19*,20*,21*,22*,23*,24*,25*} with the proper formalism created in this field. The additional poles due to $m-m$ interaction have been also introduced in the unitarized chiral perturbation theory \cite{26*,27*,28*,29*,30*,31*,32*}, see also the review paper \cite{11}. In principle these results can be obtained not using Chiral Perturbation Theory, and exploiting the dispersive methods and data, one obtains a reasonable picture of $f_0(500)$ and other scalar resonances \cite{33*,34*,35*}.

Despite of all efforts and large amount  of  information the main  problems, underlined above, were not yet fully resolved and in the PDG summary, Table 2 \cite{1} the lowest scalar resonances are identified with $f_0(1370)$ for $I=0$ and $a_0(1450)$ for $I=1$, implying that the lowest $1\,^3P_0  ~q\bar q$ pole is around (1.4-1.5) GeV, which contradicts numerous calculations in relativistic models \cite{27,28,29,30}.

In the previous paper \cite{31} the basic formalism was combined to explain the possible connection of the basic
$q\bar q$ poles to the scalar resonances $f_0(500),f_0(980)$ via the quark-chiral coefficients and meson-meson
channel-coupling interaction. In the present paper we formulate this approach in more detail, calculating all masses
and coefficients without fitting parameters, using for that the explicit form of the $q\bar q$ wave functions to calculate all
coefficients. In this way, as will be shown below, we succeed in calculating both ground state and first excited states
of all scalar mesons made of $u,d,s$ quarks.

In the present paper, as well as in the previous one \cite{31}, for theoretical formulation of the scalar meson problem the use is made of the method, similar to the non-relativistic Cornell coupled-channel mechanism \cite{32a}, developed for heavy mesons, where the pure charmonium states $c\bar c$ transform into the $D\bar D$ states  and back many times, leading  to the displacement of resulting combined  resonances. This displacement occurs via creation of a pair of light quarks and numerically is of the order or less than 50 MeV. Later on these authors have studied  displaced resonances in charmonium quantitatively  \cite{33a}, and one
of the present authors (Yu.S. together with colleagues) used the Cornell formalism  to study both charmonium and bottomonium systems \cite{34}.

The general theory of channel-coupled (CC) resonances was given in \cite{32} in a general form, not  assuming pole structures in any channel,  while the CC resonance can occur, as in the case of the $\Upsilon (nS)\pi$ system  coupled to $BB^*$ or $B^*B^*$(see last ref. in \cite{34}).  Below we are specifically interested in the $q\bar q$ poles found in relativistic path-integral formalism, coupled to a pair of chiral mesons.

One of the  basic points of this method is derivation of the transition elements between the $q\bar q$ and  the meson-meson systems, and below we use, as in \cite{31}, the chiral confining  Lagrangian (CCL) \cite{35,36,37,38}. The latter essentially uses the fact that chiral symmetry breaking (CSB) may occur not only spontaneously (without evident dynamical source), but also can be connected  with  the properties of interaction. In QCD this is the confinement property, which has the scalar property (as shown, e.g., in  recent review paper \cite{38*}). As one knows in QCD  the CSB occurs in the presence of confinement, and not proven in the deconfinement phase. Therefore the CCL, derived and introduced in \cite{35,36,37,38}, has the special form, where the confinement potential $M(r)=\sigma r$, assigned to the quark line or antiquark line, is multiplied by the standard chiral factor with chiral meson operators $U(\phi) = \exp (\phi/f \gamma_5)$. In this way both $q\bar q$ and the chiral d.o.f. are connected with known coefficients and one can immediately find the coupling coefficient, which defines the decay or transformation probability of several mesons $(1,2,3,...)$ into $q\bar q$ or vice versa. This is the fact which we shall use below and which shall enable us to find strong displacements of $\pi\pi$ and $\pi\eta$ resonances and much smaller values for $K\bar{K}$.

One can wonder whether CCL can provide the basic relations, known from the standard chiral Lagrangian SCL, e.g. the GMOR relations \cite{42}. It was shown in \cite{38} that CCL can provide two series: (1) an expansion in powers of the quark masses, with the first term yielding GMOR relations, and (2) another expansion, which yields series with powers of quark loops with derivatives of $\phi$ at the vertices, and this gives, e.g., the correct values of the terms in $O(p^4)$ \cite{38}. In this way it was shown that CCL also provides the standard and well-known chiral relations, but in addition it generates completely new relations supported by data.
Thus the formalism of CCL allows to extend the possibilities of the standard chiral formalism. As it is  the  CCL contains both chiral and the $q\bar q$ d.o.f. and this is in contrast to the standard chiral Lagrangian (SCL) and ChPT. As it was told above and will be shown later in the paper, this formalism allows to calculate all coupling constants between $q\bar q$ and two or more chiral mesons, and in particular, to calculate numerically the decay constants $f_\pi$, $f_K$, etc.\cite{42*}. As an important check of our formalism in \cite{40}, \cite{42**} and \cite{42***} the pion mass and the quark condensate  in the magnetic field were computed,  where the quark d.o.f. are essential. The results occur to be in good agreement with recent lattice data \cite{42^4} ,\cite{42^5} and \cite{42^6}, whereas the famous old  results \cite {42^7}, based on the standard chiral theory, strongly contradict those. As it is one can conclude that the extension of the famous standard chiral formalism, made in the CCL, is reasonable and can be further developed and used in QCD.

In this paper our purpose is to define the exact $q\bar q$ poles, using the detailed relativistic theory (see \cite{28,30}  and refs. therein), and establish  explicit relations between the known $^3P_0$ $q\bar q$ state characteristics and resulting  new resonance pole parameters, which will be called the Pole Projection Mechanism (PPM).

In the  framework of PPM, as shown in \cite{31}, a single $q\bar q$ pole can create one projected resonance, one  for  each meson-meson channel, coupled to a given $q\bar q$ channel. Including $\phi\phi$ channel coupling (e.g., in $\pi\pi$-$K\bar K$ channels ), one obtains two resonances connected with one $q\bar q$ pole. This mechanism was applied in the case  of the $f_0(500)$ and $f_0(980)$ resonances \cite{31}, when from the original $q\bar q$ pole with the mass $M_1=1.05$ GeV  two resonances,  $f_0(500)$ and $f_0(980)$, are created. In this way both properties, mentioned above, were demonstrated, since $f_0(500)$ occurs due to the $\pi\pi$ channel coupling to the $q\bar q$ initial state with the mass $M_1$, while $f_0(980)$ appears due to the $K\bar K$-$q\bar q$ channel coupling. Simultaneously in the case with the isospin $I=1$ and  the initial mass  $M_1$ the $q\bar q$- pole is coupled to both channels,  $\pi\eta$ and $K\bar K$, and produces two close-by  resonances near $1$ GeV, which can be
associated with $a_0(980)$. As  shown in \cite{31},  in the PPM there exists the only variable parameter -- the  spatial radius $\lambda$  of the quark-meson transition amplitude, denoted as $k^{(I)} (q\bar q,\,\varphi\varphi)$, which should be found self-consistently in our method.

The spatial radius $\lambda$ enters the quark chiral Lagrangian \cite{35,36,37,38} as the mass  parameter $M(\lambda)=\sigma \lambda$  and it is fixed in the case of $\pi, K$  mesons by  the  calculation of  the decay constants $f_\pi, f_K $ \cite{40},  which yields $\lambda =0.83$ GeV$^{-1}$. In the $q\bar q -\varphi\varphi$ transition case we calculate for the first time dependence of the coefficient $k^{(I)} (q\bar q, \varphi\varphi)$ on $\lambda$ and find a stable maximum at $\lambda=\lambda_0$ in the region $(1\leq \lambda_0\leq 1.5) $ GeV$^{-1}$, which is taken as a basic point of our method, yielding the fixed value of $k^{(I)}(\lambda_0) $ and  the  fixed $\lambda = \lambda_0$. Since $\sigma$ is known to be equal $0.18$ GeV$^2$, the meson and quark masses are fixed, $\lambda$ at the stationary point is equal $1$ GeV$^{-1}=0.20$~fm  and in this way all parameters of our formalism are fixed and known.

In present paper we further extend the PPM theory to include the radial excitations of the $q\bar q$ states and find the resulting scalar resonances. To this end we consider the $n\bar n, n\bar s, s\bar s$ states with $n_r = 0,1$ and $I=0, 1/2, 1$, and show that the inclusion of the  radial excited $q\bar q$  pole makes the PPM even more pronounced, when the lower pole, coupled with the  meson-meson  channels, has large shift down, while the second higher pole has much smaller shift. In this way we demonstrate the important visible feature of the scalar resonances: the lowest $n_r=0$ poles are much strongly shifted as compared to the $n_r=1$ poles.

To calculate the resulting shifted poles we need  1) the transition coefficients  $k^{(I)} (\lambda_0)$, discussed above; 2) the $q\bar q$ pole masses $ M_1, M_2$, computed in the framework of relativistic path integral Green's  functions \cite{43a}; and 3)  the  free $\varphi\varphi$ Green's functions $G_{\varphi\varphi} (E,\lambda_0)$, defined with the spatial distance  $\lambda_0$  between the in and out $\varphi\varphi$ states. As a result, we find  the complex energy poles,  corresponding to  observed resonances $f_0(500)$, $f_0(980)$, $f_0(1370)$, $f_0(1500)$, $a_0(980)$, $a_0(1450)$, $K_0^*(700)$, $K_0^*(1430)$ and  $f_0(1710)$.

The plan of the paper is as follows. In section 2 we present the details of the PPM formalism of \cite{31} in the case of the $I=0,1/2,1; J^P =0^+$ channels, and in section 3 we analyze  the dynamics of our theory and calculate the resulting positions of the resonances. The  inclusion of radial excited $q\bar q$ states and calculation of the resulting scalar resonances is done in section 4.  Section 5 is devoted to the discussion of results and  possible future
developments of our approach.

\section{The quark-chiral  dynamics in the  $(q\bar q)$-(meson-meson) channel)}

The main element of the Cornell formalism  \cite{32a} is the expression for the total quark-meson Green's function (resolvent) ${\cal{G}}(E)$ via the $q\bar q$ resolvent $G_{q \bar q}$ and the
meson-meson resolvent $G_{\varphi\varphi}$,

\be {\cal G} (E) = \frac{A}{1-V_{q\varphi} G_{\varphi\varphi} (E) V_{\varphi q} G_{q\bar q} (E)}, \label{1a}
\ee
so that the  resonance energies are to be  found from the equation

\be VG_{\varphi\varphi} (E) V G_{q\bar q} (E) =1, \label{2a}
\ee
where the main point is the transition element $V_{q\varphi} = V^+_{\varphi q}$.

In  \cite{32a,33a} it was shown how the channel coupling affects the charmonium poles. Later on this formalism has acquired the specific features, necessary to explain the poles in the heavy-quark systems, e.g. in $X(3872)$ \cite{34}, where the original
$2\,^3P_1$ pole of the $c\bar c$ system is strongly shifted due to transitions of $c\bar c (2^3P_1)$ into the $D\bar D^*$  meson-meson state and back, which finally provides a
pole at the $D\bar D^*$  threshold. Actually the equation for the position of the new quark-meson  pole has similar forms: nonrelativistic  \cite{32a,33a,34} and relativistic  in the new formulations for the scalars \cite{31}:

$ G_{\varphi\varphi} (E) \Gamma  G_{q\bar q} (E) \Gamma =1$, where $\Gamma$ is the $q \bar q$-$\varphi\varphi$ transition vertex, and in \cite{31} it was found that  for the chiral $\varphi\varphi$ mesons the value $\Gamma$ is large in the case of $\pi\pi$ and $\pi\eta$ systems.

Note, that  one could call $X(3872)$ as the $D \bar D^*$  resonance, but at the  same time it  can  be  considered as the shifted $c\bar c$ resonance, implying that it is the combined $c\bar c- D\bar D^*$ phenomenon, or  the $c\bar c$ pole projected on the $D\bar D^*$ channel.

At this point one realizes  that a single $c\bar c$  pole  can interact with one of $(D  \bar D, D  \bar D^*, D^*  \bar D^*) $ states and can be connected with one resonance. In the heavy-quarkonia  case the resulting  pole shifts are of the order of $\sim  50$ MeV , if the  meson-meson thresholds are nearby  the  original $Q\bar Q$ poles, whereas in the general case the situation can be different and, as shown in \cite{31},  in light mesons the pole shifts can reach 500 MeV.  At this point it is important to stress the general  features of the PPM method, when the original $(q\bar q)$ pole is projected into the meson-meson pole due to interaction between the $q\bar q$ and  the chiral meson-meson channels, implying a strong  but meson-dependent coupling. As a result, one $q\bar q$  pole can be projected originally into one  meson-meson resonance,  associated with the  corresponding meson-meson threshold, and later, taking into account the meson-meson channel coupling, can be connected with two or more resonances. As it was shown in \cite{31}, this happens in the case of the $f_0(500)$ (the $\pi\pi$ channel) and the $f_0 (980)$ (the $K\bar K$ plus coupled $\pi\pi)$, which are both connected to the $n\bar n(1^3P_0)$ pole at around 1.1 GeV.

These features create  a completely new picture of  possible ``extra  poles'', generated by the
regular  $q\bar q$ poles in QCD, not connected to any  molecular or tetraquark mechanisms. Note, that the PPM can easily be extended to  the  three-meson case $(m_1,m_2,m_3)$, coupled  to the $q\bar q$ pole, as it occurs  in the cases with the  isospin $I=1$,
$J=1,2$, namely, the $a_1(1P), a_2(1P)$ cases,  which will be discussed elsewhere.

Below we shall present the PPM, which can explain the appearance of a new pole for each new meson-meson combination, starting with  one original $q\bar q$ pole, as it was done in the $f_0(500), f_0(980)$ case. We start with the  basic element of the PPM formalism in the case of chiral mesons -- the CCL, introduced in \cite{35,36,37} and  extended  recently in \cite{38}. This Lagrangian is  a generalization  of the standard chiral theory, which takes into account  not only chiral meson but also the quark-antiquark d.o.f. The latter are necessary to calculate  the meson coupling constants $(f_\pi, f_K,...)$ \cite{42*}, to write the correct Green's functions for chiral mesons, and also  to calculate the higher $O(p^4,p^6)$ terms of chiral perturbation theory (see \cite{38}).

The CCL has the form

\be L_{CCL} =-N_c tr \log (\hat \partial+ \hat m + s_0 + \hat s + M\hat U),\label{1}
\ee
where $\hat U$ is   the  standard  chiral operator,
\be
\hat U = \exp (i \gamma_5 \hat\varphi),~~ \hat \varphi = \frac{\varphi_a \lambda_a}{f_a}, \label{2}
\ee
\be
\hat \varphi = \sqrt{2}\left(\begin{array}{ccc} \frac{1}{f_\pi} \left( \frac{\eta}{\sqrt{6}} +
\frac{\pi^0}{\sqrt{2}}\right), &\frac{\pi^+}{f_\pi},&
\frac{K^+}{f_K}\\
\frac{\pi^-}{f_\pi} & \left( \frac{\eta}{\sqrt{6}}-
\frac{\pi^0}{\sqrt{2}}\right) \frac{1}{f_\pi},& \frac{K^0}{f_{K^0}}\\
\frac{K^-}{f_K},& \frac{\bar K^0}{f_{K^0}}, & - \frac{2\eta}{\sqrt{6}
f_\pi}\end{array}\right).\label{3}\ee

Note that the CCL plays the role of the generating functional, which can produce several interesting expansions. Indeed, exploiting the trace logarithm structure of it, which allows to separate a common factor, the CCL can be transformed to the following expression \cite{38},
\be
L_{CCL} =-N_c tr \log (1- \eta),\ee
where
\be
\eta = \hat U^+ S^{-1}(\hat \partial + \hat m)(\hat U - 1),
\ee
and for $m=0$ it gives an expansion in quark loops with the quark propagators $S$, which yields $O(p^n$) terms, while the expansion in $\hat m$ to the second order yields GMOR relations. In what follows we shall not use this type of expansion, but instead we shall exploit  eq.~(\ref{3}) as it is, expanding $\hat U$ in powers of $\varphi$, keeping the second order for the meson-meson amplitude.

In eq.~(\ref{3}) $M$ is the $q\bar q$ interaction term, $M = \sigma r$, which gives confinement interaction between $q$ and $\bar q$ everywhere in the $q\bar q$ loop, however, in the vertex, where chiral mesons of $\hat U$  are  emitted, $M$ is multiplied by the operators $\varphi$. In  this  case, i.e. in the one-$\pi$, or the one-$K$, emission vertex the value of $M$, as  shown in \cite{42*}, is equal to $M(\lambda) =0.15$ GeV, which corresponds to $\lambda \cong 0.166$ fm $=0.83$ GeV$^{-1}$. In our case, when two mesons are emitted, below we shall find $\lambda$ as the stationary point of the transition coefficient, which  is equal $0.2$ fm $=$ $1$ GeV$^{-1}$.  It is interesting that it coincides with the fundamental length of the QCD vacuum, known from the Field Correlator Method (FCM) \cite{43}.

In the case of the one-meson emission vertex the  value of $M=0.15$ GeV is exactly that, which gives correctly the pion and the kaon decay constants, calculated in the framework of the CCL. From  \cite{42*} one has $$\sqrt2 f_\pi = 138 ~{\rm MeV},\,\,\sqrt2 f_K =165   ~{\rm MeV},$$ which are in good agreement with experimental values \cite{1},

 $\sqrt{f_\pi} = 130.7 \pm 0.1 \pm 0.36$ MeV, $\sqrt{f_K} = 159.8 \pm 1.4\pm 0.44$ MeV.

The important feature of the CCL is that it is directly connected to the confinement -- $\sigma r$ term -- and contains the quark d.o.f., which are absent in the standard form of the chiral Lagrangian. As was discussed in Introduction, one of immediate results of this is the correct behavior of the chiral parameters - the quark condensate, $f_\pi$, and the pion mass under the influence of the magnetic field \cite{42**} \cite{42***} as compared to recent lattice data \cite{42^4,42^5,42^6}, whereas the well-known  results of the standard chiral Lagrangian \cite{42^7} strongly contradict these data.

Note that $M(\lambda)$ is the only parameter of the CCL, which is fixed in our case (see below), in addition to the quark masses. The main idea of the quark-chiral approach \cite{35,36,37,38} is that the scalar confining operator $M(\lambda)$, violating chiral symmetry, is  augmented by the chiral operator $U(\hat \varphi)$, which can  emit any number of chiral mesons at the vertex of the $q\bar q$ operator.

Correspondingly, one can introduce the chiral-free $q\bar q$ Green's function from Eq.~(\ref{1}) with $U=1$, which we call
$G_{q\bar q}$ (see Fig.1), the free meson-meson Green's function $G_{\varphi\varphi},$ (see Fig.2), and the transition element from $q\bar q$ to  the $\varphi\varphi$ system, which is obtained from the CCL, Eq.~(\ref{1}), as shown in \cite{31} (see Fig.3).

\be \Delta L=- N_c tr \Lambda s \Lambda M(\lambda) \frac{\hat \varphi^2}{2}.\label{4}\ee

\begin{figure}[!htb]
\begin{center}
\includegraphics[angle=0,width=5 cm]{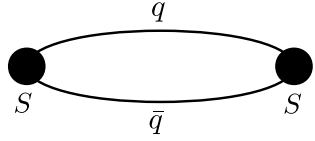}
\caption{ The scalar $q\bar q$ Green's function $G_{q\bar q}$} \vspace{1cm}
 \end{center}

\end{figure}

Here $s$ is the external current, e.g. in the $f_0(500), f_0(980)$ cases ($I=0$) it is equal to 1, while $\Lambda$ is the quark propagator, $\Lambda= (\hat \partial + m_q + M)^{-1}$.

\begin{figure}[!htb]
\begin{center}
\includegraphics[angle=0,width=7 cm]{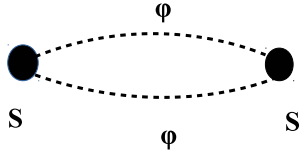}
\caption{  The scalar $\varphi\varphi$ Green's function $G_{\varphi\varphi}$} \vspace{1cm}
\end{center}
\end{figure}

At this point we can find the form of the $q \bar q$ Green's function augmented by the transition to the
$\phi\phi$ system, which is needed to start the chain of transformations, discussed here. This structure
is presented in the next figure.

\begin{figure}[!htb]
\begin{center}
\includegraphics[angle=0,width=7 cm]{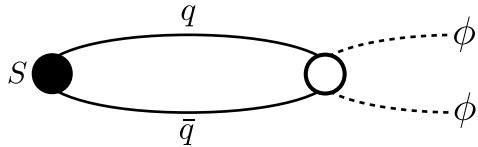}
\caption{ The scalar $q\bar q$ Green's function with the emission of the chiral
mesons} \vspace{1cm}
\end{center}
\end{figure}

As seen from (\ref{4}) and following  \cite{31}, one can find the numerical coefficient $C^{(I)}_{\varphi\varphi}$ in the transition factor $k^{(I)}(q\bar q| \varphi\varphi)$, which   defines how many ${\varphi\varphi}$ are produced by the one $q\bar q$ state. In \cite{31} this was done for isospin $I=0,1$. Here we shall consider  also the case of the $K\pi$ channel $(I=1/2$).

We conclude this section with the explicit form of the $I=1/2$ isotopic current, producing $K\pi$ in the case of the $K^*_0(700)$ resonance.

\be tr \left(j (u\bar s) \frac{\hat \varphi^2}{2}\right)= K^+ \frac{\pi^0}{\sqrt{2}} + K^0\pi^+ \label{6}\ee

\be tr \left(j (d\bar s) \frac{\hat \varphi^2}{2}\right)= K^+  {\pi^0}- \frac{\pi^0}{\sqrt{2}}   K^0 \label{7}\ee

\begin{figure}[!htb]
\begin{center}
\includegraphics[angle=0,width=5 cm]{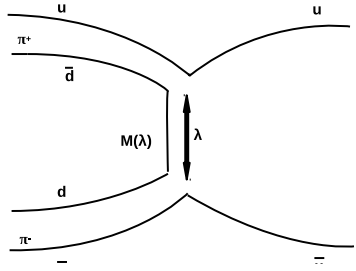}
\caption{The transition region ($q\bar q|\varphi\varphi)$ with the spatial distance $\lambda$ between the  constituents  } \vspace{1cm}
 \end{center}

\end{figure}

\section{Dynamics of the $q \bar q$ and the meson-meson systems}

The structure of the transition operator $M\frac{\hat \varphi^2}{2}$ (\ref{4}) requires a detailed investigation. In \cite{31} it was understood that the free meson-meson Green's function, created and annihilated at local points, diverges logarithmically and should be replaced by the physically motivated meson-meson Green's function, where initial and final distances between mesons are defined dynamically, i.e. by the effective distance $\lambda$ from the stationary point of the transition coefficient. In the present paper we shall follow the same line of reasoning and define the meson-meson Green's function with fixed  spatial distance $\lambda$ between the mesons at the initial and final point.

One can start with local $\varphi\varphi$ Green's function $G_{\varphi\varphi}(x,y)$, created by $\hat \varphi^2(x)$ in (\ref{4}),  $G_{\varphi\varphi}(P)$ with the total momentum $P=(E,0)$

\be G_{\varphi\varphi} (P) = \frac{1}{(2\pi)^4}\int \frac{d^4 p}{(p^2-m^2_1)((P-p)^2-m^2_2)}.\label{5}\ee

To take into account nonlocality in the initial or final vertex we shall examine the structure of this nonlocal vertex in more detail, assuming its structure as shown in Fig. 4. As seen, for the distance $\lambda$ between $q$ and $\bar q$ (and effectively between $\varphi$ and $\varphi$) one should have the corresponding Green's  functions $G_{q\bar q}$ and $G_{\varphi\varphi}$ of the form $G^{(\lambda)}_{q\bar q}(x,x'|y,y'), G^{(\lambda)}_{\varphi\varphi }( y,y'|u,u')$ with the distance $\lambda = |\vex- \vex'| \cong |\vey-\vey'|$. The  effective value of $\lambda$ in this vertex $(q\bar q|\varphi\varphi)$ is defined by the product $G^{(\lambda)}_{q\bar q} \sigma \lambda G^{(\lambda)}_{\varphi\varphi}$ which amounts to the $\lambda$ dependence of the transition coefficient and will be found below in the next sections.

The Green's function $G^{(\lambda)}_{\varphi\varphi }( y,y'|u,u')$ can be written as a product $\int \frac{d^4 p}{(2\pi)^4}
f(p) \exp(ip(y-u)) \int \frac{d^4 p'}{(2\pi)^4} f(p')\exp(ip'(y'-u')$. Now writing $\exp(i(p y + p' y') = \exp(i(p + p')(y + y')/2 + i(p-p')(y - y')/2)$ with $P=(p + p')$, one can make the Fourier transformation in $(y + y')/2$ while integrating over angles of the spatial vector $(y-y')$. The same arrangement can be done for $u,\,u'$ vectors. Denoting $p - p' = q$, one ends up with the same integral eq.(\ref{5}) , multiplied with the square of the angular integral of $\exp(i \veq \vep\lambda)$.

As it is we need the  explicit form of the meson-meson $G^{(\lambda)}_{\varphi\varphi} (y y' |u u')$  and the $q\bar q$  Green's function  $G^{(\lambda)}_{q\bar q}(x x' | y y')$, defined with the initial and final spatial   distance $\lambda$ between $\varphi$ and $\varphi$ or $q $ and $\bar q$. Since $G^{(\lambda)}_{q\bar q}$ is convergent  at $\lambda=0$, we shall consider  this effect later in this section and  now start with the effect of spatial distance $\lambda$ in $G^{(\lambda)}_{\varphi\varphi}$. As shown above, the latter amounts to the angular integration of the factors $\exp(ip(u-u'))$ and $\exp(ip(y-y'))$, where we denote $q = p$.

The result can be written in the form of the additional factor $F(\lambda p)= \left(\frac{\sin(\lambda p)}{\lambda p}\right)^2,$ $p=|\vep|$, appearing in (\ref{5}), namely,

\be  G^{(\lambda)}_{\varphi\varphi} (P) = \frac{1}{(2\pi)^4} \int \frac{d^4pF(\lambda p)}{(p^2 - m^2_1)\left((P-p)^2 - m^2_2\right)},\label{8}\ee
where $F(\lambda p) = \left( \frac{\sin (\lambda p)}{\lambda p} \right)^2, ~~ p=|\vep|$, appears due to averaging over directions of $\Delta \vey = \vey -\vey'$, $\Delta \veu=\veu-\veu'$, with $|\Delta \vey| = |\Delta \veu|=\lambda$.

The explicit form of (\ref{8}) can be written in the c.m. frame,
\ben
\operatorname{Re}G^{(\lambda)}_{\varphi\varphi} (E) = \int^\infty_0 \frac{p^2 dp}{4\pi^2} \frac{F(\lambda p)}{\sqrt{ p^2+m^2_1}\sqrt{p^2+m^2_2}}\times
\een
\be
\times\left\{ \frac{ E(\sqrt{ p^2+m^2_1}+ \sqrt{ p^2+m^2_2})+ m^2_1-m^2_2}{\left[( \sqrt{ p^2+m^2_1}+\sqrt{ p^2+m^2_2})^2-E^2\right] \left[ E + \sqrt{ p^2+m^2_1}-\sqrt{p^2+m^2_2}\right]}\right\}\label{9}\ee

\be \operatorname{Im}  G^{(\lambda)}_{\varphi\varphi} (E)= \frac{F(\lambda p_0) }{16\pi} \frac{\sqrt{ [E^2-(m_1+m_2)^2][E^2-(m_1-m_2)^2]}}{E^2}, \label{10}\ee

where $p_0$ is found from the  relation $\sqrt{ p^2_0+m^2_1}+ \sqrt{ p^2_0+m^2_2} = E \geq m_1+m$.
Another way of the  renormalization of $\oRe G_{\varphi\varphi} (E)$ was accepted in \cite{31}, with $F(\lambda p) \to 1$ and the fixed upper limit of the $p$ integration,
$p\leq N = 1/\lambda$. In what follows we shall compare both ways and find that they produce similar results.
It is clear that the factor $F(\lambda p)$ is not introduced by hand, but results from the S-wave angular integration of the product of the two-meson Green's functions at the spatial distance $\lambda$ from each other, which does not give rise to additional singularities. Note that $F(\lambda p)$ is actually a function of $\lambda^2 p^2$ and therefore it does not contribute to the difference $G_{\varphi\varphi}^{(\lambda)} (E+i\delta) - G^{(\lambda)}_{\varphi\varphi} (E -i\delta)$ on the cut $E\geq m_1+m_2$, and hence does not violate the unitarity  condition.

In the case of the $K\pi$ Green's  function one has $m_1=m_K $ (493 MeV for  $K^\pm$), and $m_2=m_\pi\cong 140$ MeV. The resulting   form  (\ref{9}) of  the $\oRe G^{(\lambda)}_{\pi K} (E)$  was computed numerically  in the range  640 MeV $\leq E \leq 1200$ MeV for $\lambda =(0.5;1;2;3)$ GeV$^{-1}$. The results of calculations show that $\oRe G^{(\lambda)}_{\pi K} (E)$ is almost constant in the  range $[0.64\div 0,9]$. For the following we shall need the values of $\oRe G^{(\lambda)}_{\pi K}$  at the point $ E = 0.64$~GeV and 0.8 GeV, given in Table~\ref{tab.01}.

{\small

\begin{table}[!htb]
\caption{The real part of the $K\pi$ Green's function as a function of the spacial distance $\lambda$ for two values of the energy,  $E=640$~MeV and$E=800$~MeV}
\begin{center}
\label{tab.01}\begin{tabular}{|c|c|c|c| }\hline

   $\lambda$ (GeV$^{-1}$    & $\oRe G^{(\lambda)}_{\pi K}$ (640 MeV)      &  $\oRe G^{(\lambda)}_{\pi K}$ (800 MeV)  &  $\oRe G^{(\lambda)}_{\pi K}$ (640 MeV, cut-off) \\\hline

0.5&0.033&0.028&0.03\\\hline

1&0.025&0.02&0.022\\\hline

1.5&0.02&0.0165&0.017\\\hline

2&0.017&0.013&0.013\\\hline

3& 0.013&0.007&\\\hline

\end{tabular}
\end{center}
\end{table}
}

In the right column of Table~\ref{tab.01} the values of $\oRe G^{(\lambda)}_{\pi K}  (E=640$ MeV) are obtained with the cut-off of the integral over $dp$ in (\ref{9}) at $N=1/\lambda$. One can see their close values, within (10-15)\%  accuracy, in
the columns 1 and 3.

Now we turn to the $q\bar q$ Green's function and shall use the same formalism for the $(n\bar s)$ system, as in \cite{31} for the  $(n\bar n)$ system; for that one can exploit calculated positions of the $(n\bar n)$ pole (see Table~\ref{tab.02}) and  analogously,   the  $(s \bar s)$, and $n\bar s$ poles. To calculate the $q\bar q$ Green's function and the $q\bar q$ eigenvalues we use,  as in \cite{31}, the exact relativistic formalism (see \cite{43a} for a review and references, based on the Field Correlator
Method \cite{43}). This yields the  relativistic Hamiltonian in the c.m. frame,  containing the quark and antiquark kinetic energies $\omega_1,\omega_2$,

\be H(\omega_1,\omega_2,\vep) = \sum_{i=1,2} \frac{\vep + \omega_i^2+m^2_i}{2\omega_i} + V_0 (r) + V_{s0}(r) + V_t\label{10*}
\ee
Now one has two options to define $\omega_i$: 1) to minimize $H(\omega_1,\omega_2,\vep)$ in the  values of $\omega_1,\omega_2$, which leads to the so-called  Spinless Salpeter Equation (SSE), widely used (see e.g. \cite{27}), or to calculate the eigenvalue of (\ref{10}) $E(\omega_1,\omega_2)$ and then to find  its minimum (so-called the ``einbein approximation'' $(EA$); see \cite{28,30,43a} for details). The comparison of these approximations for  the cases of $n\bar n$ scalar meson masses is given in Table ~\ref{tab.02}.

The interaction terms $V_0, V_{so} , V_t$ are the instantaneous  potentials of the scalar confinement $V_0$, perturbative and nonperturbative spin-orbit interactions $V_{s0}$, and tensor interaction $V_t$, which define the center-of-gravity eigenvalue $M_{\rm cog}(nP)$, the spin-orbit correction $a_{\rm so}(nP)$, and the tensor
correction $c_t(nP)$. For the masses of the $n^3P_0$ states one has \cite{28,30}

\be M(n^3P_0)=M_{\rm cog} (n^3P_0)-2a_{\rm so}-c_t. ~\label{10**}
\ee
The resulting masses of the $n\bar n$, $n\bar s, s\bar s$  states are given in the Table ~\ref{tab.02}

{\small
\begin{table}[!htb]
\caption{The masses (in MeV) of the $1^3P_0$ and $2^3P_0$ $n\bar n$ states, obtained in the SSE, EA and RT (the Regge trajectory formalism) by Badalian and Bakker \cite{28,30}, Ebert et. al. \cite{29}, and Godfrey, Isgur \cite{27} }
\begin{center}
\label{tab.02}\begin{tabular}{|c|c|c|c|c|c| }\hline
State& \multicolumn{3}{c|}{BB \cite{28,30}}&EFG \cite{29}&GI \cite{27}\\\hline
&SSE&EA&RT&&\\\hline
$n\bar n~ 1^3P_0$&1050 &1093&1038&1176&1090\\\hline
$  2^3P_0$&1461&1594&1435&1679& 1780\\\hline
\end{tabular}
\end{center}
\end{table}
}

As shown in \cite{43a,43} the $q\bar q$ Green's function can be written as a sum over the pole terms. As in \cite{31},  the lowest  pole  contribution to the $(q\bar q)$ Green's function  $G_{q\bar q} (E)$ can be written as
\be
G_{q\bar q} (E) = \sum^\infty_{n=1} \frac{(f^{(n)}_s)^2 M^2_n}{M^2_n -E^2} = \frac{(f_s^{(1)})^2 M^2_1}{M^2_1-E^2}+...\label{11}
\ee
where $f_s^{(1)}$ was calculated   in  the $(n\bar n)$ case  in \cite{31}, while for all $q\bar q$ states it is given in   Appendix A1, and within the 10\% accuracy it has the   value,   $f^{(1)}_s\cong 100$ MeV, whereas the mass $M_1(n\bar s)$ is obtained to be
$M_1= (1210\div 1240)$ MeV, and  $M_1(s\bar s)\cong 1400$ MeV, see Table ~\ref{tab.03}.

{\small
\begin{table}[!htb]
\caption{The masses (in MeV) of the $n^3P_0$ $q\bar q$ scalars, obtained in the method of \cite{28, 30}, and their experimental values in the $\pi\pi, K\bar K,\, \pi\eta,\, \pi K$ systems }
 \begin{center}
\label{tab.03}\begin{tabular}{|c|c|c|c|c|c|  }\hline

$n_r$&& $n\bar n(I=1)$& $n\bar n(I=0)$& $n\bar s \left( I=\frac12\right)$& $s\bar s (I=0)$
 \\\hline

 &$M_1$ &1.050&1050&1240&1400
 \\
 0& exp& $a_0(980)$&$f_0(980), f_0(500)$& $K^*_0(700)$& $\tilde f_0(1370)$
 \\\hline

&$M_2$ &1500&1500&1550&1740
 \\
 1&exp& $a_0(1450)$&$f_0(1500)$& $K^*_0(1430)$& $\tilde f_0(1710)$
 \\\hline

\end{tabular}

\end{center}

\end{table}
}

Now we can write the final equation for the position of the pole, resulting from the infinite  series of the $(q\bar q) \to (\varphi\varphi) \to  (q\bar q) \to ...$ transformations, in the same way as it was done in \cite{31}.

\be E^2 = M^2_1 \left\{ 1-k^{(I)} (q\bar q  |\varphi\varphi) (\oRe G^{(\lambda)}_{\varphi\varphi} (E) + i \oIm G^{(\lambda)}_{\varphi\varphi} (E))\right\},\label{12}\ee
where
\be k^{(I)} (q\bar q|\varphi_1\varphi_2) = \frac{C^2_i M^2(\lambda) (f_s^{(1)})^2}{f^2_{\varphi_1}f^2_{\varphi_2}}.\label{13}\ee

At this point it is interesting to discuss the position of the poles, which are the self-consistent solutions of the (\ref{12}). To start we consider the simplest case with equal masses of two mesons, $m_1 = m_2$ and $E^2 = p^2 + 4 m^2$, and start, solving the equation (\ref{12}) in terms of the variable $p$, taking into account that $\operatorname{Re} G$ is the constant and $\operatorname{Im} G$ is proportional to $p$, $\operatorname{Im} G =p f(p^2)$. As a result one obtains the equation for the position of the resonance in terms of $p$:
 \be p^2 +i p f(p^2) -p_0^2 = 0. \ee
As a first approximation one can take $f(p^2) = f(p_0^2) = f_0$ and solving the quadratic equation, one obtains
 \be p = -i f_0/2 +/- sqrt{ (f_0)^2/4 + p_0^2}, \ee
which explicitly shows that the pole is on the second sheet with respect to the $2m$ threshold. In next approximations one takes into account, step by step, the $p$ dependence of $f(p^2)$, observing the motion of the pole on the second sheet.

In Eq.~(\ref{13})  $C^2_i$  can be found for the $\pi\pi, K\bar K, \pi\eta$ cases as in \cite{31} and  from (\ref{6}), (\ref{7}), and  for $\pi K$ system it is equal to
\be C^2_i = \left(1+ \frac{1}{\sqrt{2}}\right)^2 = \frac32 + \sqrt2 = 2.91\approx 3, \label{14}\ee
while the PS decay constants $f_i$  are known from \cite{40}, experimental and lattice data,

\be
f_K =111~{\rm MeV}~, f_\pi = 93~{\rm MeV}~, f_\eta  =120 ~{\rm MeV}~.\label{15}
\ee
The quark decay constants of the scalar mesons $f_s^{(i)}$ are calculated via the radial derivative of the $q\bar q$ wave function, as shown in Appendix A1, with the values given in Table~\ref{tab.08}. In Appendix A2 we show that $f_s^{i)}$ are strongly dependent on the value of $\lambda$ and  the effective region of $\lambda$ is inside the range $ 0 \leq \lambda \leq 1.5$~GeV$^{-1}$. At the same  time another factor in (\ref{13}) $ M^2(\lambda)$ grows with $\lambda$, so that the optimal values of $\lambda$ can be obtained from the ratio $\frac{k^{(I)}(q\bar q|\varphi\varphi)}{k^{(I)}_{\max} (q\bar q |\varphi\varphi)}\equiv X(\lambda)$,  given
in Table~\ref{tab.04}

{\small
\begin{table}[!htb]
\caption{The  dependence of the  ratio of the transition factor $k^I(q\bar q|\varphi\varphi)/k^{(I)}_{\max}$ on the spatial contact distance $\lambda$. }
 \begin{center}
\label{tab.04}\begin{tabular}{|c|c|c|c|c| }\hline

   $\lambda$ (GeV$^{-1})$    & 0.5&1&1.5&2 \\\hline
$X(\lambda)$ &0.29&0.816&1&0.04
 \\\hline

\end{tabular}
\end{center}
\end{table}
}

Then taking into account that $M(\lambda) = \sigma \lambda =0.18$ GeV$^2\cdot \lambda$, one has  the following values of the transition factors $k^{(I)}(q\bar q|\varphi\varphi)$ at $\lambda=1$~GeV$^{-1}$ and  $\lambda=1.5$~GeV$^{-1}$ (see Table~\ref{tab.05}).

{\small
\begin{table}[!htb]
\caption{ The transition factor $k^{(I)}(q\bar q|\varphi\varphi)$ at $\lambda =1$~GeV$^{-1}$ and
$\lambda=1.5$~GeV$^{-1}$ for different channels}
 \begin{center}
\label{tab.05}\begin{tabular}{|c|c|c|c|c|c| }\hline

   $k(q\bar q|\varphi\varphi)$  & $(n\bar n|\pi\pi)$ & $(n\bar n| KK)$& $(n\bar n |\pi\eta)$& $(n\bar s | \pi K)$& $(s\bar s | KK)$  \\\hline

$\lambda =1$ GeV$^{-1}$  &18.44&4.02&3.0& 14.2&3.0
 \\\hline

$\lambda =1.5$ GeV$^{-1}$  &41.51&9.05&6.72& 31.2&6.75
 \\\hline

\end{tabular}
\end{center}
\end{table}
}

Using these values of  $k^{(I)}(\bar q\bar q|\varphi\varphi)$ in Eq.~(\ref{12})  and the values of $M_1$ from Table~\ref{tab.03},   one obtains the parameters of the resonances in the  channels $\pi\pi, KK, \pi\eta $ $\pi K$, given in Table~\ref{tab.06}.

{\tiny
\begin{table}[!htb]
\caption{ The resonances in the channels $\pi\pi, K\bar K, \pi\eta, \pi K$, coupled at the distance 1 GeV$^{-1}$ and $\lambda=1.5$ GeV$^{-1}$ to the $q\bar q$ poles  $(n\bar n, n\bar s, s\bar s)$,  in comparison with experimental PDG data }
 \begin{center}
\label{tab.06}\begin{tabular}{|c|c|c|c|c|c|c| }\hline

&
   $ (q\bar q|\varphi\varphi)$  & $(n\bar n|\pi\pi)$ & $(n\bar n| K\bar K)$& $(n\bar n |\pi\eta)$& $(n\bar s | \pi K)$& $(s\bar s | K\bar K)$  \\\hline
&$ k(q\bar q|\varphi\varphi)$&18.44&4.02&3.0 &14.2&3.0\\ \cline{2-7}
$\lambda=1$& $\oRe G_{\varphi\varphi}$& 0.02&0.011&0.02&0.025&0.011\\\cline{2-7}
 &$\oIm G_{\varphi\varphi}$&0.015&0.02&0.015&0.015&0.02\\\cline{2-7}
 & $\oRe a, \oIm a$& 0.38;0.276&0.045+i0.08&0.06+i0.045&0.36+
 i0.213&0.033+i0.06\\\cline{2-7}
 &$E$&0.85-i0.17&1.025-i0.044&1.02-i0.025&0.714-i0.078&1.37-i0.041\\
  \hline

&$ k(q\bar q|\varphi\varphi)$&41.51&9.05&6.75 &31.2&6.75\\ \cline{2-7}

$\lambda=1.5$& $\oRe G_{\varphi\varphi}$& 0.015&0.018&0.018&0.0165&0.018\\\cline{2-7}
 &$\oIm G_{\varphi\varphi}$&0.0155&0.015&0.015&0.015&0.015\\\cline{2-7}

 & $\oRe a, \oIm a$& 0.645;0.645&0.162+i0.136&0.1215+i0.10&0.52+
 i0.468&0.12+i0.10\\\cline{2-7}

 &$E$&0.64-i0.54&0.966-i0.08&0.98-i0.056&0.75-i0.21&1.31-i0.074\\ \cline{2-7}
 &$E_{PDG}$&0.400-0.550&0.990&0.980&0.630-0.730&1.200-1.500\\\cline{2-7}
 &$\Gamma_{PDG}$&0.400-0.700&0.010-0.100&0.050-0.100&
 0.478(50)&0.200-0.500\\
  \hline
\end{tabular}
\end{center}
\end{table}
}
From Table~\ref{tab.06} one can see that suggested the pole projection mechanism (PPM) yields a reasonable picture of the resulting resonances in all $\varphi\varphi$ channels, and the  differences between calculated and observed resonance characteristics $(R,\Gamma)$ are of the order of indeterminacy intervals. A possible sign of disagreement seems to be in the $f_0(500)$ resonance, where PPM gives a  resonance position some 150-200 MeV  above the experimental value. As it was discussed in \cite{31}, this fact implies that the $\pi\pi$ interaction in the $\pi\pi$ Green's function, $G_{\pi\pi}(E)$, has to be used to account for the low energy region, $E\la 500$ MeV. Indeed, the accurate analysis in \cite{45} confirms the   $f_0(500)$ pole position at $E=(457-i 279)$ MeV, close to $E_{\exp}$. If this interaction is neglected, from Table~\ref{tab.06} for $\lambda=(1,1.5)$ GeV$^{-1}$ we have

 \be  E({\rm GeV})=(0.85\div 0.64)- i (0.17\div 0.54),  \label{20}\ee
which   differs from $E_{exp},$ while the $f_0(980)$ data is comparable to  our result.

\section{ The  case of two $q\bar q$ poles }

Till now we have studied the lowest $^3P_0$  quark-antiquark poles, which due to the PPM are shifted down from the original position of around (1000$-1400)$ MeV to the final position in the range (700-1300) MeV, which  can be associated with the  lowest exotic   resonances. However,  in the ($n\bar n)$ channel there is  the radially excited pole $0^{++}, I=0$ at the initial position $M_1=(1490-1500)$ MeV, which can be also shifted down and have the  position around 1400~MeV, known as $f_0(1500)$. Also in the $K_0^*$-channel ($J^{PC}=0^{++}, I=\frac12$)  there exists the higher resonance, coupled to the  same $K \pi$ decay  channel, $K_0^*(1430)$, which can be originated from the radial excited $(n\bar s)$ pole at $M_2=1550$ MeV. Below we shall show  a remarkable property of  the PPM, where the shift down of the lowest $(q\bar q)$ pole changes a little, if the radial excitations are taken into account, while the mass shift of the higher $(q\bar q)$ pole is strongly suppressed as compared to the ground state. This property of the level repulsion follows from the structure of the PPM equations themselves.

Indeed, writing the one-channel, one-pole PPM Eq.({35}) in the form  as in \cite{31},  one has
\be G_{\varphi\varphi}(E)   k^{(I)}(q\bar q|\varphi\varphi)\frac{M^2_1}{M^2_1-E^2}=1,  \label{23}\ee
with
  \be k^{(I)} (q\bar q|\varphi\varphi) = \frac{(C_{\varphi\varphi}^{(I)})^2 M^2(\lambda) (f^{(1)}_s)^2}{f_\varphi^4}, ~~ f_\varphi= f_\pi,   f_K, f_\eta
\label{24}\ee

This equation can be generalized, including the radially excited pole  $M_2,$ as follows
\be G_{\varphi\varphi}(E)  \left[ k^{(I)}_1(q\bar q|\varphi\varphi)\frac{M^2_1}{M^2_1 -E^2}+k^{(I)}_2(q\bar q|\varphi\varphi)\frac{M^2_2}{M^2_2 -E^2}\right]=1
\label{25}\ee

\begin{center}

\begin{figure}[ht]
\begin{center}
  \includegraphics[width=8cm, ]
  {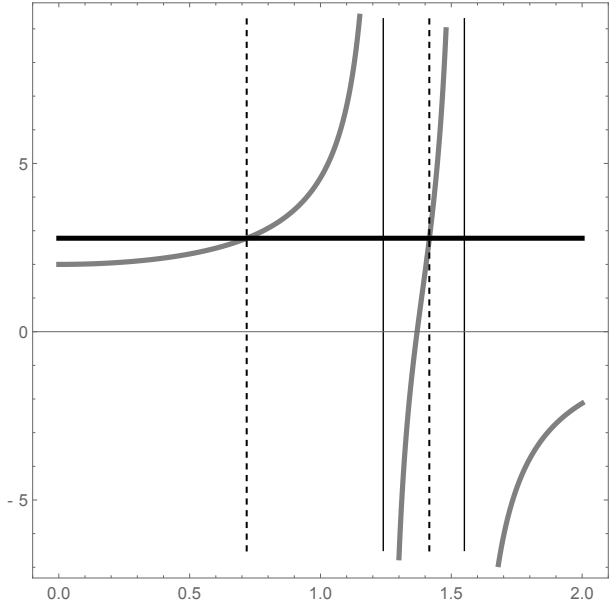}

\end{center}
  \caption{The function $f(E)$ is shown by thick  grey  lines,   with two poles at $E=M_1, M_2$ (shown by  thin  vertical lines). The intersections of $f(E)$ with the horizontal   line at $a^{-1}=1/0.36$  yields  two resulting  poles  $E=E_1, E_2$, marked by  vertical dashed lines}

\end{figure}
 \end{center}

To understand better the situation with two projected poles we consider  the Eq.(\ref{25}) and  approximate $k_1^{(I)}\approx k_2^{(I)} $ (which  holds in most cases according to Table ~\ref{tab.08} in Appendix A1). From (\ref{25}) one has the equation

  \be f(E) = \frac{M^2_1}{M^2_1-E^2}+ \frac{M^2_2}{M^2_2-E^2}= \frac{1}{k^{(I)}G_{\varphi\varphi} (E)}= a^{-1}, \label{36}\ee
Then  taking  the case $(n\bar s|\pi K)$ as an example  and neglecting $\oIm G_{\varphi\varphi}$,  from Table~\ref{tab.06}  one obtains $a=k^{(1/2)} \oRe G_{\pi K}=0.36$, and the resulting $f(E)$, as a function of $E$, has  two poles, defined by the intersection  of the straight line $f(E) = \frac{1}{0.36}$ (see Fig. 5). From Fig. 5 one can easily see how the resulting poles $E_1,E_2$ are shifted as  compared to $M_1, M_2$, in the approximation of zero $\oIm G_{\pi K}$.

  { To proceed with  the case of $K^*_0(700), K^*_0(1430)$,} we are
solving the quadratic in $E^2$  equation (\ref{36}) with $M_1=1.24, ~~ M_2=1.55$ GeV,    and  obtain  two  approximate solutions for $\lambda =1$~ GeV$^{-1}$
  \be E_1 = (0.78 - i~0.33)~{\rm GeV}, ~~ E_2 =(1.40-i~0.035)~{\rm GeV}. \label{37}\ee

These solutions correspond to the  intersection points in Fig.5 and were obtained treating the imaginary part of $G_{\pi K} (E)$   as perturbation. To take it fully into account one can write the solution of (\ref{36}) as

\be E^2=\frac12 (M^2_1+M^2_2) (1-a) \pm \sqrt{\frac14 ( M^2_1+M^2_2)^2 (1-a)^2-  M^2_1 M^2_2  (1-2a)}\label{38}\ee
and use   $$  a = \oRe a + i\oIm a = k^{(I)} (q\bar q | \varphi\varphi) (\oRe G_{\varphi\varphi}+ i\oIm G_{\varphi\varphi}) $$  from the Table~\ref{tab.06} in the  case $(n\bar s|\pi K)$, calculated e.g.  for $\lambda=1$ GeV$^{-1}$.

In a similar way one can consider all the cases:  $(n\bar n|\pi\pi),$ $ (n\bar n| K\bar K),$ $ (n\bar n|\pi\eta),$ $ (n\bar s | K \pi)$ and $(s\bar s| K\bar K)$. The resulting pole positions for $\lambda=1$ GeV$^{-1}$, generated by ground and radially excited scalar $q\bar q$ poles, are  given in the Table~\ref{tab.07}.

{\tiny
\begin{table}[!htb]
\caption{ The scalar resonance positions and the widths in the two-pole formalism}
 \begin{center}
\label{tab.07}\begin{tabular}{|c|c|c|c|c|c| }\hline
 The   $ (q\bar q|\varphi\varphi)$  connection  & $(n\bar n|\pi\pi)$ & $(n\bar n| K\bar K)$& $(n\bar n |\pi\eta)$& $(n\bar s | \pi K)$& $(s\bar s | K\bar K)$  \\\hline
$n_r=0$& 1.05&1.05&1.05&1.24&1.4\\
The $q\bar q$ mass (GeV)&&&&&\\\cline{2-6}
$n_1=1$&1.50&1.5&1.5&1.55&174\\&&&&&\\\hline
Transition coefficient &18.44&4.02&3.0&14.2&3.0 \\ $k^{(I)}(q\bar q|\varphi\varphi )$ &&&&& \\\hline
$a(E)=k^{(I)} G_{\varphi\varphi}(E)$&0.38+i0.28&0.045+i0.08&0.06+i0.045&0.36+i0.213&0.033+i0.06\\\hline
$E_1(n_r=0)$ (GeV), &0.8 &1.04 &1.02 &0.85 &1.36\\\cline{2-6}
$\Gamma_1$(MeV)&980&32&40&640& 72 \\\hline

&$f_0(500)$&$f_0(980)$&$a_0(980)$&$K_0^*(700)$&$f_0(1370)$?\\\hline

$E^{(1)}_{PDG} $ (GeV)&0.40-0.55&0.99&0.98&0.63-0.73&1.2 $\div 1.5 $\\\cline{2-6}

$\Gamma_{PDG}$ (MeV) &400-700&0.10-100&0.50-100&480&200$\div$500\\\hline

$E_2(n_r=1)$(GeV) &1.28 &1.45 &1.45 &1.4 &1.72\\\cline{2-6}

$\Gamma $ (MeV)& 100&84&52&40& 76\\\hline

&$f_0(1370)$&$f_0(1500)$&$a_0(1450)$&$K_0^*(1430)$&$f_0(1710)$\\\hline

$E^{(2)}_{PDG} $&1200-1500&1.50&1.48&1.425&1.72\\\cline{2-6}

$\Gamma $ (MeV) &200$\div$500&$\Gamma=109$&$\Gamma=265$&$\Gamma=270$&$\Gamma=120$\\\hline

\end{tabular}
\end{center}
\end{table}
}

 From Table~\ref{tab.07} one can see a reasonable agreement of predicted and observed resonance characteristics, but  with a few exclusions. The first one refers  to the higher position of the predicted mass $f_0(500)$ with $E_1=800$ MeV, however, with a large width, which implies significant uncertainty  in the resonance position, and, as we discussed  above, calls for the account of the $\pi\pi$ interaction in $G_{\pi\pi}$ at small energies.  The second discrepancy might be more significant. Namely, the first $(s\bar s| K\bar K)$ resonance occurs exactly at 1.37 GeV (see Table~\ref{tab.07}) and could be associated with $f_0(1370)$, however, the latter prefers to decay into $\pi\pi, 4\pi$ and the $K\bar K$ ratio is less than 10\% \cite{1}.

 At the same time the second $(n\bar n|\pi\pi)$ resonance is predicted at around 1.3 GeV with the width $\Gamma_{\pi\pi}\approx  100$~ MeV, and the $(n\bar n | K\bar K)$ resonance is at 1.45 GeV with the width $\Gamma_{K\bar K }\approx 100$ MeV; the latter has to be associated with  $f_0(1500)$. Unfortunately $f_0(1500)$ decays mostly into $\pi\pi, 4\pi$. Thus one faces three  inconsistencies  in the theory: $\pi\pi$ resonance at 1300 MeV and two $K\bar K$ resonances at 1450 MeV and  1360 MeV, while in experiment one has two   resonances $f_0(1370)$ and $f_0(1500)$, decaying mostly into $\pi\pi$ and $4\pi$.

 Evidently, here appears a strong mixing pattern of  three (or more) resonances, which can be additionally enlarged by  the code mechanism $(K\bar K|n\bar n)\frac{M_2}{M^*_2-E^2}(n\bar n|\pi\pi)$ near the $n\bar n$ pole at $M_2=1.5$ GeV. As an additional argument for this mixing and the resulting damping of the $K\bar K$ decay mode, one can use the small value of the $K\bar K$ decay width of 70 MeV for the $(s\bar s| K\bar K)$ resonance at 1.36 GeV, while corresponding experimental resonance $f_0(1370)$ has a large $\pi\pi, 4\pi$ width, $\Gamma =(200\div 500)$ MeV.   This interesting topic requires a substantial  analysis and a separate publication.

\section{Conclusions and  an outlook}

In our paper we presented the simplest version of the  channel coupling (CC) mechanism with the code -- $(q\bar q|\varphi\varphi)$, which is the relativistic and the chiral extension of  original the Cornell code, used for the charmonium  resonances \cite{32a}. This is the realization of  the CC mechanism \cite{32}, where due to  infinite set of transformations of one system into another one can provide a pole (the bound state) in this  set,  even if both systems are free. Here the basic role is played by the magnitude of the transition amplitude and  the concrete  example  of the resulting $Z_b$ resonances was given in the last refs. of \cite{34}.

It was also demonstrated that in the case of scalar mesons the role of transition coefficient $k^{(I)}(q\bar q|\varphi\varphi)$ is extremely important, since it can be very large number, $(k^{(0)}(n\bar n|\pi\pi) =O(18-40))$ in the $(n\bar n|\pi\pi)$  and the $(n\bar s|K\pi)$ cases (see Table~\ref{tab.06}),  and small, $(k=O(1))$, in other cases. In Tables~\ref{tab.06} and \ref{tab.07} one can see that just this large range of the changes helps to understand  the situation with the scalar mesons, where the shifts of the resonances are so different in different $\varphi\varphi$ systems, and the maximal one is in the  $(n\bar n|\pi\pi)$ case.

At this point one can see the main difference of the present approach from other existing formalisms. As was explained in the paper,  the connection between $q\bar q$ and the meson-meson channels plays the basic role and starting from the single $q\bar q$ pole, one can define the parameters of all lowest scalar resonances (this does not mean that other mechanisms are ruled out). However, here one must find the transition coefficients explicitly, without fitting parameters, which we could do with the use of the CCL and the stationary point in the function $k^{(I)}(\lambda)$. An approximate way of adjusting this connection was already used in the unitarized meson model \cite{19*,20*,21*}, where one needs to introduce two to three parameters to describe the transitions. Another approach is the dispersive method \cite{45}, where the rigorous integral equations are used in comparison with data. At this point we should stress that the final adjustment of the positions of the lowest scalar resonances, obtained in our formalism without fitting parameters, nevertheless, requires the use of these results, as was demonstrated in \cite{31}, where the data of \cite{45} were essentially used.

Another important feature of the PPM is the  appearance of the resonance, created by a single  $q\bar q$ pole -- the resulting $\varphi\varphi$ resonance can appear, in principle, in each $\varphi\varphi$ system, connected to this $q\bar q$ pole. To have more resonances, connected to the same $q\bar q$ pole, one needs additional direct $\phi_1\phi_1$-$\phi_2\phi_2$ interaction. This happens for $\pi\pi$ and $K\bar K$ systems, where two resonances $f_0(500)$ and $f_0(980)$ are created in this way by the $q\bar q$ pole at $E=1050$ MeV.  Note, that  finally these resonances become connected due to the  $\pi\pi-K\bar K$ channel coupling, and in some cases two close-by resonance poles can be located on different sheets, as was observed in  lattice analysis by J.~Dudek et al.  \cite{18}.

We have already stressed the important role of the $\varphi\varphi$ interaction in obtaining the correct position of lowest resonances $f_0(500)$ and $K^*_0(700)$. Actually our approach provides an  alternative way for the  description of the $\varphi\varphi$ scattering amplitudes, when the $q\bar q$ dynamics is included at the  first stage, and the $q\bar q-\varphi\varphi$ transition is taken into account as a second step, and the final stage should include  the detailed  account of the $\varphi\varphi$ interaction.  The  comparison of the resulting $\pi\pi$ amplitude, using only two first steps, with the realistic $\pi\pi$ data, done in \cite{31}, exactly shows that the two-step  amplitude roughly describes  main  features - the extrema and zeros of the amplitude, but strongly distorts  the amplitude at small energies, where the $\varphi\varphi$ interaction is important.  To solve the scalar meson problem, as it was demonstrated above, the simplified two-step procedure was sufficient. On another hand, the full three-step procedure provides the exact $\varphi\varphi$  amplitude with the  correct $q\bar q$ input, as it was shown in \cite{31}.

Another feature of the PPM, found in this paper, is the relatively smaller shifts of all radial excited resonances, compared to the ground states, especially in the $(n\bar n |\pi\pi)$ and $(n\bar s|\pi K)$ cases. As a whole, we have explained the general features of the scalar meson spectrum, leaving the details of the  $K\bar K-\pi\pi$ coupling to the future publications.

The  work of two of the authors (M.L. and Yu.S.) is supported  by the Russian Science Foundation in the framework of the scientific project, Grant 16-12-10414.


\vspace{1cm}

\section*{Appendix A1. Decay constants of the $n\bar n$, $n\bar s$ and $s\bar s$ states}

 \setcounter{equation}{0} \def\theequation{A1.\arabic{equation}}

As it was explained in \cite{31}, the $q\bar q$ Green's function is  computed in the Fock-Schwinger formalism, based on the relativistic path integral method. In this formalism the $q\bar q$ Green's function in the c.m. frame $(\veP =0)$ has the form

\be G_{q\bar q} (E) = \sum_n  \frac{(f_s^{(n)})^2M^2_n}{M^2_n -E^2} \to \frac{(f_s^{(1)})^2 M^2_1}{M^2_1-E^2} +
\frac{(f_s^{(2)})^2 M^2_2}{M^2_2-E^2},\label{a1}
\ee
where $M_n, n=1,2,$ are the energy eigenvalues, while $f_s^{(n)}$ are the $P$-wave decay constants, which are discussed and calculated in the  Appendix 1 of \cite{31}.

Here we only  detalize the  explicit form  of $f_3^{(1)}$ and its dependence on the quark masses and the radial quantum number $n$.
The explicit form of $f_s^{(n)}$ can be writen as \cite{31}

\be (f_s^{(n)})^2 = \frac{2N_c(R'_{nP}(0))^2}{4\pi \omega_n \bar \omega_n  M_n}, \label{a2}
\ee
where $\omega_n, \bar \omega_n$ are the average energies of the quark and the antiquark in the relativistic $q\bar q$ system obeyed by the confinement, the color Coulomb and spin-dependent interactions \cite{30}. The concrete calculations, done in this framework as in \cite{31}, bring the following results presented in the Table~\ref{tab.08}.

\begin{table}[!htb]
\caption{The quark kinetic energy $\omega_i$ (in GeV), the derivative of the radial wave function at the origin $R'_{iP}(0)$, the masse $M_i$ (in GeV), and the decay constant $f_s^{(i)}$ for the ground state ($i=1$) and the first excited state ($i=2$)}
 \begin{center}
\label{tab.08}\begin{tabular}{|c|c|c|c|c|c|c|c| }\hline

$q\bar q$& $\omega_1;\omega_2$ & $R'_{1P}(0)$ & $R'_{2P}(0)$ & $M_1$ & $M_2$ &$(f_s^{(1)})^2$ & $(f_s^{(2)})^2$ \\ && (GeV$^{5/2}$) & (GeV$^{5/2}$) & & & (GeV$^2$) & (GeV$^2$) \\\hline
$n\bar n$ & 0.48; 0.50&0.0845&0.0906&1.05&1.5&0.0142&0.0103
 \\\hline

$n\bar s$ & 0.53; 0.56&0.091&0.106&1.24&1.55$\div$1.61&0.010&0.0108
 \\\hline

$s\bar s$ & 0.54; 0.57 &0.099&0.116&1.4&1.74&0.0112&0.0101
 \\\hline

\end{tabular}

\end{center}
\end{table}

\section*{Appendix A2.}

\setcounter{equation}{0} \def\theequation{A2.\arabic{equation}}
As it is shown in (\ref{a2}), the decay constant $f_s^{(n)}$ ($s$ - the scalar) is defined via the derivative $R'_{nP} (0)$, while other factors in (\ref{a2}) do not depend on $r$.

For the decay constant, defined at the spatial distance $r=\lambda$  between $q$ and $\bar q$ (see Fig. 4), the decay constant $f_s^{(n)}(\lambda)$ is determined via the derivative $R'_{nP}(\lambda)$, i.e. generalizing Eq.(\ref{a2}),
\be (f_s^{(n)} (\lambda))^2 = \frac{2N_c (R'_{nP}(\lambda))^2}{4\pi\omega_n \bar \omega_n M_n}\label{A2.1}\ee

The values of $R'_{nP}(\lambda)$ have been computed numerically in the relativistic formalism of \cite{28,30}  and  corresponding values of $R'_{1P}(\lambda), (R'_{1P}(\lambda))^2$ are given in the Table~\ref{tab.09} together with the ratios of the decay constants $\eta(\lambda) =\left|\frac{f_s(\lambda)}{f_s(0)}\right|^2$

\begin{table}[!htb]
\caption{The space distances $\lambda$, the derivative of the wave function $R'_{1P}(\lambda)$ and $(R'_{1P}(\lambda))^2$, and the parameter $\eta(\lambda)$ for the ground $n\bar n$ state}
 \begin{center}
\label{tab.09}\begin{tabular}{|c|c|c|c|c|c|c|c|c| }\hline

$\lambda$ (GeV$^{-1}$)& 0.25& 0.50&0.75&1.0&1.25&1.50&1.75&2.0  \\\hline

$R'_{nP}(\lambda)$ (GeV$^{5/2}$)&  0.0852&0.082&0.0764&0.0684&
 0.06&0.0504&0.0101&0.0077
 \\\hline

$(R'_{nP}(\lambda))^2$ GeV$^5$ &
0.00726&0.00672&0.00583&0.00468
&0.0036&0.0025&0.0001&5.9$\cdot 10^{-5}$
 \\\hline

$\eta(\lambda) = \left|\frac{f_s^{(1)}(\lambda)}{f_s^{(1)}(0)}\right|^2$
  &  0.98&0.91&0.79&0.63&0.486&0.343&0.0138&0.008\\
  \hline

\end{tabular}

\end{center}
\end{table}

\end{document}